\documentclass[10pt,a4paper,onecolumn]{article}
\usepackage[a4paper, total={6.5in, 10.4in}]{geometry}
\usepackage{setspace,xurl}
\usepackage{lineno,hyperref,multirow,cite}
\usepackage{longtable}
\usepackage{authblk}
\usepackage{epsfig,amssymb,amsmath}
\usepackage[utf8]{inputenc}
\usepackage{amsmath,graphicx}
\usepackage{amssymb,amsmath,epsfig,url,mathrsfs} \usepackage{algorithm,algorithmic}
\usepackage[usenames, dvipsnames]{color}
\usepackage[framemethod=TikZ]{mdframed}
\usepackage{xcolor}
\usepackage{colortbl}
\usepackage[colorinlistoftodos]{todonotes}
\usepackage{multirow}
\usepackage{multicol}
\usepackage{enumitem}
\usepackage{hhline}

\usepackage{numprint}
\npthousandsep{,}

\usepackage{graphicx}
\usepackage{subfig}
\usepackage{array, makecell}
\usepackage{verbatim}
\usepackage[utf8]{inputenc}
\usepackage[T1]{fontenc}
\usepackage[ngerman,english]{babel}
\usepackage{tabularx}  % for 'tabularx' environment and 'X' column type
\usepackage{ragged2e}  % for '\RaggedRight' macro (allows hyphenation)
\newcolumntype{Y}{>{\RaggedRight\arraybackslash}X} 
\usepackage{booktabs}  % for \toprule, \midrule, and \bottomrule macros 
\usepackage{amsthm}
 
\usepackage{tcolorbox}
\usepackage{graphicx}
 
\definecolor{chestnut}{rgb}{0.8, 0.36, 0.36}
\definecolor{frenchblue}{rgb}{0.0, 0.45, 0.73}
\definecolor{babyblue}{rgb}{0.54, 0.81, 0.94}
\definecolor{beaublue}{rgb}{0.74, 0.83, 0.9}
\definecolor{bubblegum}{rgb}{0.99, 0.76, 0.8}
\definecolor{gainsboro}{rgb}{0.86, 0.86, 0.86}
\definecolor{amethyst}{rgb}{0.6, 0.4, 0.8}
\definecolor{mygray}{gray}{0.85}
\definecolor{aq}{rgb}{0.5, 1.0, 0.83}
\definecolor{aquamarine}{rgb}{0.5, 1.0, 0.83}
\definecolor{beaublue}{rgb}{0.74, 0.83, 0.9}
\definecolor{cadetblue}{rgb}{0.37, 0.62, 0.63}
\definecolor{lavender}{rgb}{0.9, 0.9, 0.98}
\definecolor{nonphotoblue}{rgb}{0.64, 0.87, 0.93}
\definecolor{paleaqua}{rgb}{0.74, 0.83, 0.9}
\definecolor{peachpuff}{rgb}{1.0, 0.85, 0.73}
\definecolor{uscgold}{rgb}{1.0, 0.8, 0.0}
\definecolor{pastelorange}{rgb}{1.0, 0.7, 0.28}
\definecolor{mountainmeadow}{rgb}{0.19, 0.73, 0.56}
\definecolor{mossgreen}{rgb}{0.68, 0.87, 0.68}
\definecolor{lemonchiffon}{rgb}{1.0, 0.98, 0.8}
\definecolor{lightblue}{rgb}{0.68, 0.85, 0.9}
\definecolor{azuremist}{rgb}{0.94, 1.0, 1.0}
\definecolor{caribbeangreen}{rgb}{0.0, 0.8, 0.6}
\definecolor{celadon}{rgb}{0.67, 0.88, 0.69}
\definecolor{camouflagegreen}{rgb}{0.47, 0.53, 0.42}
\definecolor{coolgrey}{rgb}{0.55, 0.57, 0.67}
\definecolor{darkolivegreen}{rgb}{0.33, 0.42, 0.18}
\definecolor{darkspringgreen}{rgb}{0.09, 0.45, 0.27}
\makeatletter
\newcommand{\mathleft}{\@fleqntrue\@mathmargin0pt}
\newcommand{\mathcenter}{\@fleqnfalse}
\makeatother
 
\theoremstyle{definition}

\usepackage{todonotes}

\hypersetup{
   unicode=false,          % non-Latin characters in Acrobat’s bookmarks
   pdftoolbar=true,        % show Acrobat’s toolbar?
   pdfmenubar=true,        % show Acrobat’s menu?
   pdffitwindow=false,     % window fit to page when opened
   pdfstartview={FitH},    % fits the width of the page to the window
   pdftitle={VoicePrivacy 2024 Challenge Evaluation Plan},    % title
   pdfauthor={VoicePrivacy},     % author
   pdfsubject={VoicePrivacy},   % subject of the document
   pdfcreator={},   % creator of the document
   pdfproducer={}, % producer of the document
   pdfkeywords={VoicePrivacy, anonymisation}, % list of keywords
   pdfnewwindow=true,      % links in new window
   colorlinks=true,       % false: boxed links; true: colored links
   linkcolor=blue,          % color of internal links
   citecolor=blue,        % color of links to bibliography
   filecolor=magenta,      % color of file links
   urlcolor=blue           % color of external links
}

\usepackage{color, colortbl}
\definecolor{LightCyan}{rgb}{0.88,1,1}
\definecolor{cambridgeblue}{rgb}{0.64, 0.76, 0.68}
\definecolor{cambridgebluemy}{rgb}{0.74, 0.86, 0.78}
\definecolor{cambridgebluemy2}{rgb}{0.94, 0.99, 0.92}
\definecolor{beige}{rgb}{0.96, 0.96, 0.86}
\mdfdefinestyle{MyFrame}{%
    linecolor=black,
    outerlinewidth=.1pt,
    roundcorner=1pt,
    innertopmargin=\baselineskip,
    innerbottommargin=\baselineskip,
    innerrightmargin=5pt,
    innerleftmargin=-1pt,
    backgroundcolor=gray!10!white}

\title{The VoicePrivacy 2024 Challenge\\ Evaluation Plan\\[1em]\large{}Version \colorbox{cambridgebluemy2}{\textbf{2.0}}}

\author[1]{Natalia Tomashenko}
\author[2]{Xiaoxiao Miao}
\author[1]{Pierre Champion}
\author[3]{Sarina Meyer}
\author[4]{Xin Wang}
\author[1]{Emmanuel Vincent}
\author[5]{Michele Panariello}
\author[5]{Nicholas Evans}
\author[4]{Junichi Yamagishi}
\author[5]{Massimiliano Todisco}

\affil[1]{Inria, France}
\affil[2]{Singapore Institute of Technology, Singapore}
\affil[3]{Institute for Natural Language Processing, University of Stuttgart, Germany}
\affil[4]{National Institute of Informatics, Tokyo, Japan}
\affil[5]{Audio Security and Privacy Group, EURECOM, France}

\date{\url{https://www.voiceprivacychallenge.org/}}

\begin{document}

\maketitle
% \vspace{-4pt}
% organisers@lists.voiceprivacychallenge.org

\small
\begin{tcolorbox}[width=\textwidth, colback={cambridgebluemy2}, title={\textbf{For new participants --- Executive summary}}, colbacktitle=cambridgebluemy, coltitle=black, arc=0.3mm, fonttitle=\bfseries, boxrule=0.5pt]    
  \begin{itemize}[leftmargin=3.5mm]\setlength\itemsep{0.0em}
      \item The task is to develop a voice anonymization system for speech data which 
      conceals the speaker's voice identity while protecting linguistic content and 
      emotional states.
      \item The organizers provide development and evaluation datasets and evaluation scripts, as well as baseline anonymization systems and a list of training resources formed on the basis of the participants' requests. Participants apply their developed anonymization systems, run evaluation scripts and submit evaluation results and anonymized speech data to the organizers.
      \item Results will be presented at a workshop held in conjunction with  
      Interspeech 2024
      to which all participants are invited to 
      present their challenge systems and to
      submit additional workshop papers.
  \end{itemize}
\end{tcolorbox}

\vspace{3pt}

\begin{tcolorbox}[width=\textwidth, colback={cambridgebluemy2}, title={\textbf{For readers familiar with the VoicePrivacy Challenge --- Changes w.r.t.\ 2022}}, colbacktitle=cambridgebluemy, coltitle=black, arc=0.3mm, boxrule=0.5pt]    
   \begin{itemize}[leftmargin=3.5mm]\setlength\itemsep{0.0em}
    
     \item In line with the considered application scenarios, the requirements that anonymization preserves voice distinctiveness and intonation are removed, hence the associated $G_{\text{VD}}$ and $\rho^{F_0}$ metrics are no longer used.
     All the data are anonymized on the \textit{utterance level}.

     \item An extended list of datasets and pretrained models, formed on the basis of the participants' requests, will be provided for training anonymization systems.

     \item The complexity of the evaluation protocol and the running time of the evaluation scripts have been greatly reduced. The scripts are now primarily in Python, which makes it easy for participants who are new to the field to catch up. 

     \item Only objective evaluation will be performed. Three complementary metrics will be used: the equal error rate (EER) as the privacy metric and two utility metrics, namely the word error rate (WER) for automatic speech recognition (ASR) and the unweighted average recall (UAR) for speech emotion recognition (SER).

     \item Models for utility evaluation (ASR and SER) are trained on original (unprocessed) data to ensure that linguistic and emotional content is undistorted.
     These models are provided with the evaluation scripts, hence utility evaluation is much faster.

   \end{itemize}
\end{tcolorbox}

\vspace{3pt}

\begin{tcolorbox}[width=\textwidth, colback={cambridgebluemy2}, title={\textbf{Changes in version 2.0 w.r.t. 1.0}}, colbacktitle=cambridgebluemy, coltitle=black, arc=0.3mm, boxrule=0.5pt]    
   \begin{itemize}[leftmargin=3.5mm]\setlength\itemsep{0.0em}

    \item The final list of data and models to build and train  anonymization systems (Table~\ref{tab:data-models-final-list}).
    
     \item New anonymization baselines: \textbf{B3}, \textbf{B4}, \textbf{B5}, and \textbf{B6} (Sections~\ref{sec:baseline_b3}, \ref{sec:baseline_b4}, \ref{sec:baseline_b5_b6})
     and 
     %corresponding 
     results (Section~\ref{subsec:results}).

   \end{itemize}
\end{tcolorbox}

\newpage

\normalsize

\section{Challenge objectives}
Speech data fall within the scope of privacy regulations such as the European General Data Protection Regulation (GDPR). Indeed, they encapsulate a wealth of personal (a.k.a.\ personally identifiable) information such as the speaker's identity, age and gender, health status, personality, racial or ethnic origin, geographical background, social identity, and socio-economic status \cite{Nautsch-PreservingPrivacySpeech-CSL-2019}.
Formed in 2020, the VoicePrivacy initiative \cite{tomashenko2020introducing} is spearheading efforts to develop privacy preservation solutions for speech technology. So far, it has focused on promoting the development of \emph{anonymization} solutions which conceal all personal information, facilitating their comparison using common datasets and protocols, and defining meaningful evaluation metrics through a series of competitive benchmarking challenges. The first two editions of VoicePrivacy were held in 2020 and 2022 \cite{tomashenko2020introducing,Tomashenko2021CSl,Tomashenko2021CSlsupplementay, tomashenkovoiceprivacy, tomashenko2022voiceprivacy,results2022}.
VoicePrivacy 2024, the third edition, starts in March 2024 and culminates in the VoicePrivacy Challenge workshop held in conjunction with the 4nd  Symposium on Security and Privacy in Speech Communication (SPSC)\footnote{\label{fn:spsc}4th Symposium on Security and Privacy in Speech Communication: \url{http://www.spsc2024.mobileds.de/}}, a joint 
event co-located with Interspeech 2024\footnote{\url{https://www.interspeech2024.org/}} in Kos Island, Greece.

%SPSC
Anonymization requires a combination of solutions to alter not only the speaker's voice, but also linguistic content, extra-linguistic traits, and background sounds which
might reveal the speaker's identity.
In keeping with the previous VoicePrivacy Challenge editions, the current edition focuses on the subgoal of \textit{voice anonymization}, that is the task of altering the speaker's voice to conceal their identity to the greatest possible extent, while leaving the linguistic content and paralinguistic attributes intact. Specifically, this edition focuses on preserving the emotional state, that is the key paralinguistic attribute in many real-world application scenarios of voice anonymization, e.g., in call centers to enable the use of third-party speech analytics. In the following, we often refer to ``voice anonymization'' as ``anonymization'' alone for the sake of conciseness.

This document describes the challenge task, the data, pretrained models and baseline systems that participants can use to build their own voice anonymization system, and the evaluation metrics and rules that will be used for assessment, in addition to guidelines for registration and submission.

\section{Task}
\label{sec:task}

Privacy protection is formulated as a game between a \emph{user} who 
shares data for a desired downstream task and an \emph{attacker} who accesses this data or data derived from it and uses it to infer information about the data subjects \cite{qian2018towards,srivastava2019evaluating,tomashenko2020introducing}. 
Here, we consider the scenario where the user shares anonymized utterances for downstream automatic speech recognition (ASR) and speech emotion recognition (SER) tasks, and the attacker wants to identify the speakers from their anonymized utterances.

\subsection{Voice anonymization task}
\label{subsec:user_goals}

The utterances shared by the user are referred to as \emph{trial} utterances. In order to hide the identity of the speaker within each utterance, the user passes the utterance through a voice anonymization system prior to sharing. The resulting utterance sounds as if it was uttered by another speaker, which we refer to as a \emph{pseudo-speaker}. The pseudo-speaker might, for instance, be an artificial voice not corresponding to any real speaker.

The task of challenge participants is to develop this voice anonymization system. 
It should: 
\begin{enumerate}[label=(\alph*)]
    \item output a speech waveform; 
    \item conceal the speaker identity on the \emph{utterance level}; 
    \item not distort the linguistic and emotional content.

\end{enumerate}

The utterance-level anonymization requirement (b) means that the voice anonymization system must assign a pseudo-speaker to each utterance independently of the other utterances. The pseudo-speaker assignment process must be identical across all utterances and not rely on speaker labels. When this process involves a random number generator, the random number(s) generated must be different for each utterance, typically resulting in a different pseudo-speaker for each utterance. Voice anonymization systems that assign a single pseudo-speaker to all utterances also satisfy this requirement.

The achievement of requirement (c) is assessed via \emph{utility} metrics.
Specifically, we will measure the WER and UAR obtained by ASR and SER systems trained on original (unprocessed) data.

\subsection{Attack model}
\label{subsec:attack_model}

For each speaker of interest, the attacker is assumed to have access to utterances spoken by that speaker, which are referred to as \textit{enrollment} utterances. He then uses an automatic speaker verification (ASV) system to re-identify the speaker corresponding to each anonymized trial utterance.

In this work, we assume that the attacker has access to:
\begin{enumerate}[label=(\alph*)]
\item several enrollment utterances for each speaker;
\item the voice anonymization system employed by the user.
\end{enumerate}
Using this information, the attacker anonymizes the enrollment utterances to reduce the mismatch with the trial utterances, and trains an ASV system adapted to that specific anonymization system. This attack model is the strongest known to date, hence we consider it as the most reliable for privacy assessment.

The protection of identity information is assessed via a \emph{privacy} metric. Specifically, we will measure the EER obtained by the attacker.

\section{Data and pretrained models}\label{sec:data}
Publicly available resources will be used for the training, development and evaluation of voice anonymization systems. The development and evaluation data are fixed, while the choice of training resources is open to the participants.

\subsection{Training resources}
In addition to the training data used in the previous challenge editions and those used to train the baseline anonymization systems (see Section~\ref{sec:baseline}), the participants were allowed to propose additional resources to build and train anonymization systems before the deadline (20th March). These include both data and pretrained models.
Based on the  suggestions received from the challenge participants, in this version of the evaluation plan, we publish the final list of training data and pretrained   models  allowed for training anonymization systems. All the allowed resources are listed in Table~\ref{tab:data-models-final-list}.

For models $\#$ 1, 2, 3, 4, 5, 7, and 8, the provided link is a webpage listing multiple versions of the model. In this case, unless otherwise stated, all model versions available on that page before 21st March 2024 can be used by participants in the development and training of their anonymization systems.
% \item
% 2.
%
Participants are allowed to use any existing software in the development and training of their anonymization systems. If the software uses pretrained models, these models should be explicitly listed in this table. This includes models $\#$ 32-35 and the models listed on the main page (readme) of the repository $\#$ 36 before 21st March 2024.

For the purpose of the challenge, the \textit{MSP-Podcast} \cite{lotfian2017building}  corpus providers can share the \textit{MSP-Podcast}  corpus for companies using the academic license. If a company wants to use the corpus beyond this challenge, it will have to obtain a commercial license by approaching the corpus providers.

\begin{longtable}[htbp!]{|c|p{3.6cm}|p{11.3cm}|}
%\centering
  \caption{Final list of models and data for training anonymization systems.}\label{tab:data-models-final-list}\\
% \resizebox{\textwidth}{!}{
%  \centering
%  \begin{tabular}{|c|p{5cm}|p{10cm}|}
\Xhline{0.7pt}
$ \textbf{\#}$	&	\textbf{Model}	&		\textbf{Link}	\\ \hhline{===}   
1	&	WavLM  Base~and~Large \cite{Chen2021WavLM}	&	\url{	https://github.com/microsoft/unilm/tree/master/wavlm	} \\ \hline 
2	&	Whisper \cite{radford2023robust}	&	\url{	https://github.com/openai/whisper	} \\ \hline
3	&	HuBERT \cite{hubert} 	&	\url{	https://github.com/facebookresearch/fairseq/blob/main/examples/hubert	} \\ \hline
	% &		&	\url{		} \\
4	&	XLS-R \cite{babu2021xls}	&	\url{	https://github.com/facebookresearch/fairseq/blob/main/examples/wav2vec/xlsr	} \\ \hline
5	&	wav2vec 2.0 \cite{baevski2020wav2vec}	&	\url{	https://github.com/facebookresearch/fairseq/tree/main/examples/wav2vec	} \\
	&		&	\url{	https://dl.fbaipublicfiles.com/voxpopuli/models/wav2vec2_large_west_germanic_v2.pt	} \\ \hline
6	&	wav2vec2-large-robust-12-ft-emotion-msp-dim \cite{wagner2023dawn}	&	\url{	https://huggingface.co/audeering/wav2vec2-large-robust-12-ft-emotion-msp-dim	} \\ \hline
7	&	ContentVec \cite{qian2022contentvec}	&	\url{	https://github.com/auspicious3000/contentvec	} \\ \hline
8	&	w2v-BERT \cite{chung2021w2v}	&	\url{	https://github.com/facebookresearch/fairseq/tree/ust/examples/w2vbert	} \\ \hline
9	&	ECAPA2 \cite{thienpondt2023ecapa2}	&	\url{	https://huggingface.co/Jenthe/ECAPA2	} \\ \hline
10	&	ECAPA-TDNN	\cite{desplanques2020ecapa} &	\url{	https://huggingface.co/speechbrain/spkrec-ecapa-voxceleb	} \\ \hline
11	&	NaturalSpeech 3	\cite{ju2024naturalspeech} &	\url{	https://huggingface.co/amphion/naturalspeech3_facodec	} \\ \hline
12	&	\makecell[l]{NVIDIA~Hifi-GAN \\  Vocoder (en-US) \cite{kong2020hifi}} &	\url{	https://huggingface.co/nvidia/tts_hifigan	} \\ \hline
13	&	CRDNN on CommonVoice 14.0 English 	&	\url{	https://huggingface.co/speechbrain/asr-crdnn-commonvoice-14-en	} \\ \hline
14	&	Encodec \cite{encodec}	&	\url{	https://huggingface.co/facebook/encodec_24khz	} \\ \hline
15	&	Bark	&	\url{	https://huggingface.co/suno/bark	} \\
	&		&	\url{	https://huggingface.co/erogol/bark/tree/main	} \\ \hline 
\multicolumn{3}{l}{}  \\ \hline

$ \textbf{\#}$		&	\textbf{Dataset}	&	\textbf{Link}	\\ \hhline{===}

16	&	ESD \cite{zhou2021seen}	&	\url{ 	https://hltsingapore.github.io/ESD/download.html} \\ \hline
17	&	LibriSpeech \cite{panayotov2015librispeech}: train-clean-100, train-clean-360, train-other-500	&	\url{	https://www.openslr.org/12	} \\ \hline
18	&	CREMA-D \cite{cao2014crema}	&	\url{	https://github.com/CheyneyComputerScience/CREMA-D} \\ \hline
19	&	RAVDESS \cite{livingstone2018ryerson}	&	\url{	https://datasets.activeloop.ai/docs/ml/datasets/ravdess-dataset/} \\
	&		&	\url{	https://zenodo.org/records/1188976} \\ \hline
20	&	VCTK \cite{yamagishi2019cstr}	&	\url{	https://datashare.ed.ac.uk/handle/10283/2651} \\ 
	&		&	\url{	https://huggingface.co/datasets/vctk} \\ \hline
21	&	SAVEE \cite{haq2009speaker}	&	\url{	http://kahlan.eps.surrey.ac.uk/savee/} \\
& & \url{https://www.kaggle.com/datasets/ejlok1/surrey-audiovisual-expressed-emotion-savee}	 \\ \hline
22	&	EMO-DB	\cite{burkhardt2005database}&	\url{http://emodb.bilderbar.info/download/} \\ \hline
23	&	LJSpeech \cite{ljspeech17} 	&	\url{https://keithito.com/LJ-Speech-Dataset/} \\ \hline
24	&	Libri-light \cite{kahn2020libri} (only train part)	&	\url{https://github.com/facebookresearch/libri-light/blob/main/data_preparation/README.md} \\ \hline
25	&	VoxCeleb-1,2 \cite{chung2018voxceleb2}	& \url{https://www.robots.ox.ac.uk/~vgg/data/voxceleb/index.html\#about} \\ \hline
26	&	LibriTTS \cite{zen2019libritts}: train-clean-100,  train-clean-360, train-other-500	&	\url{	https://openslr.org/60/	} \\ \hline
27	&	CMU-MOSEI	\cite{zadeh2018multimodal} &	\url{	http://multicomp.cs.cmu.edu/resources/cmu-mosei-dataset/	} \\ \hline
28	&	MUSAN \cite{snyder2015musan}	&	\url{	https://www.openslr.org/17/	} \\ \hline
29	&	RIR 	\cite{ko2017study} &	\url{	https://www.openslr.org/28/	} \\ \hline
30	&	VGAF \cite{sharma2021audio} (from &	\url{	https://sites.google.com/view/emotiw2023} \\ 
	&	 EmotiW	challenge)	&	\url{https://www.kaggle.com/datasets/amirabdrahimov/vgaf-dataset} \\ \hline
31	& MSP-Podcast \cite{lotfian2017building} & \url{https://ecs.utdallas.edu/research/researchlabs/msp-lab/MSP-Podcast.html} \\ \hline
\multicolumn{3}{l}{}  \\ \hline

 $\textbf{\#}$		&	\textbf{Software with pretrained models}	&	\textbf{Link}	 \\ \hhline{===}
32	&	Resemblyzer	&	\url{	https://github.com/resemble-ai/Resemblyzer	} \\
	&		& 	Model: 	\url{https://github.com/resemble-ai/Resemblyzer/blob/master/resemblyzer/pretrained.pt	} \\ \hline
33	&	VITS \cite{kim2021conditional}	&	\url{	https://github.com/jaywalnut310/vits/	} \\ 
	&		&	Models: \url{	 https://drive.google.com/drive/folders/1ksarh-cJf3F5eKJjLVWY0X1j1qsQqiS2	} \\ \hline
34	&	PIPER pretrained on 	&	\url{	https://github.com/rhasspy/piper/?tab=readme-ov-file	} \\ 
	&	VITS	&	Models:  \url{	https://huggingface.co/datasets/rhasspy/piper-checkpoints/tree/main	} \\ \hline
35	&	RVC-Project	&	\url{	https://github.com/RVC-Project	} \\
	&		&	Models: \url{	 https://huggingface.co/lj1995/VoiceConversionWebUI/tree/main	} \\ \hline
36	&	DISSC \cite{maimon2022speaking}	&	\url{	https://github.com/gallilmaimon/DISSC	} \\ \Xhline{0.7pt}

%\end{tabular}
%}
\end{longtable}
\normalsize

\subsection{Development and evaluation data}\label{subsec:dev_eval_data}
Development and evaluation data
comprise subsets of the following corpora:
\begin{itemize}
    \item \textit{\textbf{LibriSpeech}}\footnote{\label{fn:url1}LibriSpeech: \url{http://www.openslr.org/12}} \cite{panayotov2015librispeech} is a corpus of read English speech derived from audiobooks and designed for
ASR research. It contains \numprint{960}~hours of speech sampled at 16~kHz. This data will be used for ASV and ASR evaluation.
The \textit{LibriSpeech} evaluation and development sets are the same as in the previous challenge editions.

   \item \textit{\textbf{IEMOCAP}} \cite{Busso08} is an emotional audio-visual dataset that will be used for SER evaluation.
It contains 12 hours of speech sampled at 16~kHz corresponding to improvised and scripted two-speaker conversations between 5 female and 5 male English actors.
We consider only 4 emotions out of the 9 annotated ones: \textit{neutral}, \textit{sadness}, \textit{anger}, and \textit{happiness}.
Following \cite{Pappagari2020XVectorsME,asr_IEMOCAP,Nourtel_spsc_emotion}, we merge the original happiness and excitement classes into the happiness class to balance the number of utterances in each class.
To accommodate for the small number of speakers and the small amount of data, we adopt a leave-one-conversation out cross-validation protocol. In each cross-validation fold, four conversations (eight speakers) are used to train the SER evaluation system\footnote{Trained SER evaluation systems corresponding to the 5 folds are provided by the organizers. The participants should not use this data for their own training purposes.}, while the two speakers from the remaining conversation form the development and evaluation sets, respectively.

\end{itemize}

A detailed description of the datasets provided for  development and evaluation is presented in Tables~\ref{tab:data} and \ref{tab:data_imo} below.

\begin{table}[!htbp]
\centering
  \caption{Statistics of the \textit{LibriSpeech} development and evaluation sets for ASV and ASR evaluation.}\label{tab:data}
% \resizebox{0.95\textwidth}{!}{
  \centering
  \begin{tabular}{|c|l|l|r|r|r|r|}
\Xhline{0.7pt}
 \multicolumn{3}{|l|}{\textbf{Subset}} &  \textbf{Female} & \textbf{Male} & \textbf{Total} & \textbf{\#Utterances}  \\ \hline \hline
% devel
\multirow{2}{*}{\rotatebox{0}{~ Development }} & LibriSpeech & Enrollment & 15 & 14 & 29 & 343\\ \cline{3-7}
& dev-clean & Trial & 20 & 20 & 40 & \numprint{1978}\\ \cline{1-7}
\multirow{2}{*}{\rotatebox{0}{Evaluation~}} & LibriSpeech & Enrollment & 16 & 13 & 29 & 438\\ \cline{3-7}
& test-clean & Trial & 20 & 20 & 40 & \numprint{1496}\\ \Xhline{0.7pt}
\end{tabular}%}
\end{table}
\normalsize

\begin{table}[!htbp]
\centering
  \caption{Construction and statistics of the \textit{IEMOCAP}  development and evaluation sets for SER evaluation. \textit{Train} subsets refer to the training data for the SER evaluation system.
  }\label{tab:data_imo}
 % \resizebox{0.95\textwidth}{!}{
  \centering
  \begin{tabular}{|c|l|c|c|c|c|c|c|}
\Xhline{0.7pt}
 \multicolumn{2}{|l|}{\textbf{Conversation}}  & \textbf{\#Utterances}  & \textbf{Fold 1} & \textbf{Fold 2} & \textbf{Fold 3} & \textbf{Fold 4} & \textbf{Fold 5}\\ \hline \hline

\multirow{2}{*}{\rotatebox{0}{~ Session 1 }}  &  Female  & 528 & Dev &  \multirow{2}{*}{\rotatebox{0}{{Train}}} &  \multirow{4}{*}{\rotatebox{0}{{Train}}} &  \multirow{6}{*}{\rotatebox{0}{{Train}}} &  \multirow{8}{*}{\rotatebox{0}{{Train}}} \\\cline{2-4}
 & Male  & 557 & Eval &&&&\\ \cline{1-5}
\multirow{2}{*}{\rotatebox{0}{~ Session 2 }}  &  Female  & 481 & \multirow{8}{*}{\rotatebox{0}{{Train}}} & Eval &&&\\ \cline{2-3}\cline{5-5}
 & Male  & 542 & & Dev &&&\\ \cline{1-3}\cline{5-5}\cline{6-6}
 \multirow{2}{*}{\rotatebox{0}{~ Session 3 }}  &  Female  & 522 & & \multirow{6}{*}{\rotatebox{0}{{Train}}} & Dev&&\\ \cline{2-3}\cline{6-6}
 & Male  & 629&&  &Eval&&\\ \cline{1-3}\cline{7-7}\cline{6-6}
  \multirow{2}{*}{\rotatebox{0}{~ Session 4 }}  &  Female  & 528&&  &\multirow{4}{*}{\rotatebox{0}{{Train}}} &Eval&\\ \cline{2-3} \cline{7-7}
 & Male  & 503&&  &&Dev&\\ \cline{1-3} \cline{7-7}\cline{8-8}
   \multirow{2}{*}{\rotatebox{0}{~ Session 5 }}  &  Female  & 590&&  &&\multirow{2}{*}{\rotatebox{0}{{Train}}}&Eval\\ \cline{2-3}\cline{8-8}
 & Male  & 651&& &&&Dev\\ \cline{1-3}\cline{8-8}
 \Xhline{0.7pt}
\end{tabular}
% }
\end{table}

\section{Privacy and utility evaluation}
\label{sec:perf}

We consider one objective 
privacy metric to assess the speaker re-identification risk and two objective
utility metrics to assess the fulfillment of the downstream tasks specified in Section~\ref{sec:task}.

\subsection{Objective assessment of the privacy-utility tradeoff}
\label{sec:perf_objective}

Three metrics will be used for the objective ranking of submitted systems: the equal error rate (EER) as the privacy metric and two utility metrics: word error rate (WER) and unweighted average recall (UAR).
These metrics rely on automatic speaker verification (ASV),  automatic speech recognition (ASR), and speech emotion recognition (SER) systems.
The ASV system is trained on \textit{LibriSpeech-train-clean-360} and the ASR system on the full \textit{LibriSpeech-train-960} dataset, whose statistics are presented in Table~\ref{tab:train-eval-metrics}. The SER system for each \textit{IEMOCAP} cross-validation fold is trained on the corresponding \textit{IEMOCAP} training subset, whose statistics are reported in Table~\ref{tab:data_imo}.
Training and evaluation will be performed with the provided recipes and models.\footnote{Evaluation scripts: \url{https://github.com/Voice-Privacy-Challenge/Voice-Privacy-Challenge-2024}} 
More specifically, models for privacy evaluation will be trained by participants on their anonymized training data with the provided training scripts, while the models for utility evaluation are provided by the organizers.

\begin{table}[!htbp]
  \caption{Statistics of the \textit{LibriSpeech} training sets for ASV and ASR evaluation.}\label{tab:train-eval-metrics}
  \renewcommand{\arraystretch}{1.1}
  \centering
  \begin{tabular}{|c|c|c|r|r|r|r|}
\Xhline{0.7pt}
  \multirow{2}{*}{\textbf{System}} & \multirow{2}{*}{\textbf{Subset}} &   \multirow{2}{*}{\textbf{Size,h}} & \multicolumn{3}{c|}{\textbf{\#Speakers}} &  \multirow{2}{*}{\textbf{\#Utterances}} \\ \cline{4-6}
&  &  & \textbf{Female} & \textbf{Male} & \textbf{Total} & \\ \hline \hline
 %$ASV_\text{eval}^\text{anon}$ 
 ASV
 & LibriSpeech train-clean-360 & 363.6 & 439 & 482	 &	921	& \numprint{104014}	\\ \hline
 ASR
% $ASR_\text{eval}$ 
 & LibriSpeech train-960 & 960.9 & 1128 & 1210	 &	2338	& \numprint{281241}	\\ \Xhline{0.7pt}
  \end{tabular}
\end{table}

As in the 2022 edition, multiple evaluation conditions specified with a set of minimum target privacy requirements will be considered.
For each minimum target privacy requirement, submissions that meet this requirement will be ranked according to the resulting utility for each utility metric separately. 
The goal is to measure the privacy-utility trade-off at multiple operating points, e.g.\ when systems are configured to offer better privacy at the cost of utility and vice versa.  This approach to assessment aligns better the VoicePrivacy Challenge with the user expectation of privacy and allows for a more comprehensive evaluation of each solution, while also providing participants with a set of clear optimisation criteria.  The privacy and  utility metrics will be used for this purpose.

Minimum target privacy requirements  
are specified with a set of $N$ minimum target EERs: \{EER$_1$, \ldots, EER$_N$\}. 
Each minimum target EER constitutes a separate evaluation condition.
Participants are encouraged to submit solutions to as many conditions as possible. Submissions to any one condition $i$ should achieve an average EER on the VoicePrivacy 2024 evaluation set greater than the corresponding EER$_i$.
The set of valid submissions for each EER$_i$ will then be ranked according to the corresponding WER and UAR. The VoicePrivacy 2024 Challenge involves $N=4$ conditions with minimum target EERs of: EER$_1=10\%$, EER$_2=20\%$,
EER$_3=30\%$, EER$_4=40\%$.

\begin{figure}[!htbp]
\centering\includegraphics[width=165mm]{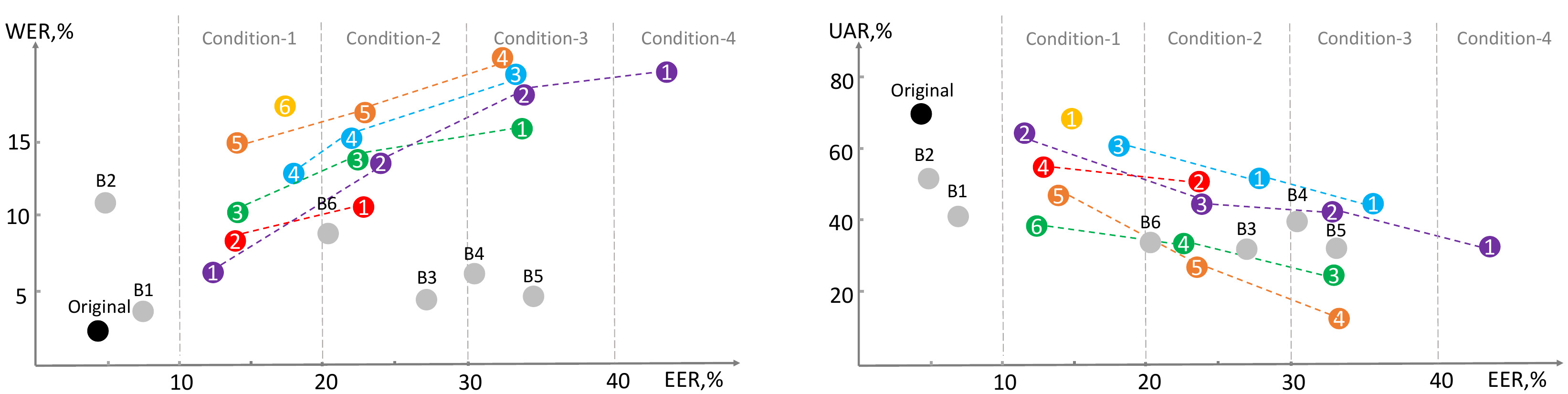}
\caption{Example system rankings according to the privacy (EER) and utility (WER and UAR) results for 4 minimum target EERs. Different colors correspond to 6 different teams. Numbers within each circle show system ranks for a given condition. Grey circles correspond to the baseline systems, and the black circle to the original (unprocessed) data.}
\label{fig:thresholds}
\end{figure}

The lower the WER 
for a given EER condition, the better the rank of the considered system in ASR results ranking. Similarly, the higher the UAR for a given EER condition, the better the rank of the considered system in SER results ranking.
A depiction of example results and system rankings according to this methodology is shown in Figure~\ref{fig:thresholds}.

\subsection{Privacy metric: equal error rate (EER)}\label{sec:asv-eval}

\begin{figure}[!htbp]
\centering\includegraphics[trim=0 280 0 0,clip,width=150mm]{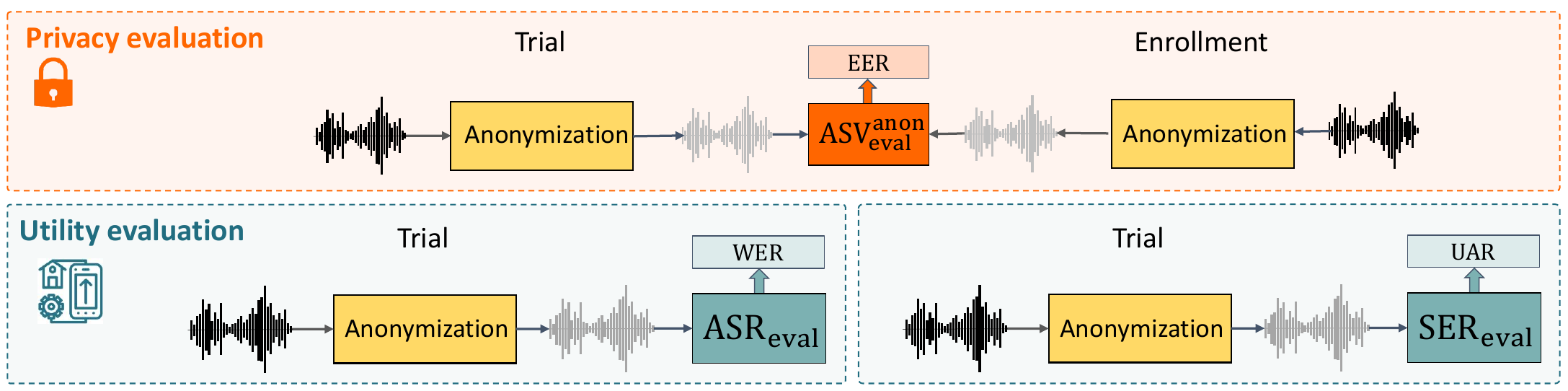}
\caption{Privacy and utility evaluation.
}
\label{fig:eval}
\end{figure}

The ASV system used for privacy evaluation is an ECAPA-TDNN \cite{desplanques2020ecapa} with 512 channels in the convolution frame layers, implemented by adapting the \textit{SpeechBrain} \cite{speechbrain} \textit{VoxCeleb} recipe to \textit{LibriSpeech}.
As seen in Figure~\ref{fig:eval}, we consider a \textit{semi-informed} attacker, who
has access to the anonymization system under evaluation \cite{srivastava2019evaluating,Tomashenko2021CSl}. Using that system, the attacker anonymizes the original enrollment data so as to reduce the mismatch with the anonymized trial data.
In addition, the attacker anonymizes the \textit{LibriSpeech-train-clean-360} dataset and retrains the ASV system (denoted $ASV_\text{eval}^{\text{anon}}$) on it, so that it is adapted to this specific anonymization system.\footnote{It is critical that the $ASV_\text{eval}^{\text{anon}}$ system is well trained, indeed a badly trained system can overestimate the EER and give a false sense of privacy \cite{champion2023}. The organizers will use the anonymized data submitted by the participants to check it. In the event when some submissions do not satisfy it, the organizers reserve the right to modify the ASV evaluation scripts or to mark those submissions accordingly to ensure a fair competition.} Anonymization is conducted on the \textit{utterance level}, using the same pseudo-speaker assignment process as the trial data.  
For a given speaker, all enrollment utterances 
are used to compute an average 
speaker vector
for enrollment.

\begin{table}[!htbp]
  \caption{Number of same-speaker and different-speaker pairs considered for evaluation.}\label{tab:trials}
  \centering
   % \resizebox{0.87\textwidth}{!}
   {
  \begin{tabular}{|l|l|l|r|r|r|}
\hline
 \multicolumn{2}{|l|}{\textbf{Subset}} & \textbf{Trials} &  \textbf{Female} & \textbf{Male} & \textbf{Total}  \\ \hline \hline
% dev
\multirow{2}{*}{\rotatebox{0}{Development~}} & LibriSpeech & Same-speaker & 704 & 644 & \numprint{1348} \\ \cline{3-6}
 & dev-clean & Different-speaker	& \numprint{14566} & \numprint{12796} &	\numprint{27362} \\ \cline{1-6}
\multirow{2}{*}{\rotatebox{0}{Evaluation~}} & LibriSpeech & Same-speaker & 548 & 449	& \numprint{997} \\ \cline{3-6}
  & test-clean & Different-speaker & \numprint{11196} & \numprint{9457} &	\numprint{20653} \\ \cline{1-6}
  \end{tabular}}
\end{table}
\normalsize

For every pair of enrollment and trial speaker vectors in the \textit{LibriSpeech} development and evaluation sets, the cosine similarity score is computed from which a same-speaker vs.\ different-speaker decision is made by thresholding.
Denoting by $P_\text{fa}(\theta)$ and $P_\text{miss}(\theta)$ the false alarm and miss rates at threshold~$\theta$, the EER metric corresponds to the threshold $\theta_\text{EER}$ at which the two detection error rates are equal, i.e., $\text{EER}=P_\text{fa}(\theta_\text{EER})=P_\text{miss}(\theta_\text{EER})$. The higher the EER, the greater the privacy.
The number of same-speaker and different-speaker pairs is given in Table~\ref{tab:trials}.

\subsection{Utility metrics}

\subsubsection{Word error rate (WER)}\label{sec:wer}

The ability of the anonymization system to leave the linguistic content undistorted is assessed using an ASR system\footnote{\url{https://huggingface.co/speechbrain/asr-wav2vec2-librispeech}} (denoted $ASR_\text{eval}$) fine-tuned on \textit{LibriSpeech-train-960} from \textit{wav2vec2-large-960h-lv60-self}\footnote{\url{https://huggingface.co/facebook/wav2vec2-large-960h-lv60-self}}
using a \textit{SpeechBrain} recipe.
Unlike the 2022 challenge edition, this ASR evaluation model is fixed, and trained and fine-tuned on original (unprocessed) data.

For every anonymized trial utterance in the \textit{LibriSpeech} development and evaluation sets, the ASR system outputs a word sequence. The WER is calculated as
\begin{equation*}
\text{WER}=\frac{N_\text{sub}+N_\text{del}+N_\text{ins}}{N_\text{ref}},
\end{equation*}
where $N_\text{sub}$, $N_\text{del}$, and $N_\text{ins}$ are the number of substitution, deletion, and insertion errors, respectively, and $N_\text{ref}$ is the number of words in the reference. %
The lower the WER, the greater the utility.

\subsubsection{Unweighted average recall (UAR)}
The ability of the anonymization system to leave the emotional content undistorted is assessed using an SER system (denoted $SER_\text{eval}$) trained using the \textit{SpeechBrain} recipe for SER on \textit{IEMOCAP}. It is a wav2vec2-based model
that has been trained separately for each of the training folds in Table~\ref{tab:data_imo}.

Within each fold, emotion recognition performance is quantified on the anonymized \textit{IEMOCAP} development and evaluation sets using the standard UAR metric calculated as  
the sum of class-wise recalls $R_i$ divided by the number of classes $N_\text{class}$:
\begin{equation*}
    \label{eq_uar}
   \text{UAR} = \frac{\sum_{i=1}^{N_\text{class}}{R_{i}}}{N_\text{class}}.
\end{equation*}
The recall $R_i$ for each class $i$ is computed as number of true positives divided by the total number of samples in that class.
The obtained UARs are then averaged across the five folds.
The higher the UAR, the greater the utility.

\section{Baseline voice anonymization systems}\label{sec:baseline} 

Baseline voice anonymization systems are released to help participants develop their own system.
We provide a description and evaluation results for two established baseline systems inspired from past challenge editions, that will be used to gauge progress with respect to these editions (\textbf{B1} and \textbf{B2}). In addition, we also have released several new baseline systems (\textbf{B3}, \textbf{B4}, \textbf{B5}, and \textbf{B6}) that better protect privacy and have different performance in utility. Note that the training data for the new baseline systems may differ for each method. All the data and models used in their development can be used by the challenge participants in the training of their anonymization system. These data and models are included in the list of training resources (see Section \ref{sec:data}).

\subsection{Anonymization using x-vectors and a neural source-filter model: B1}

The baseline anonymization system \textbf{B1} is based on a common approach to x-vector modification and speech synthesis. It is identical to the \textbf{B1.b} baseline from the VoicePrivacy 2022 Challenge \cite{tomashenko2022voiceprivacy}, except that anonymization is now performed on the utterance level instead of the speaker level.

\begin{figure}[h!]
\centering\includegraphics[width=110mm]{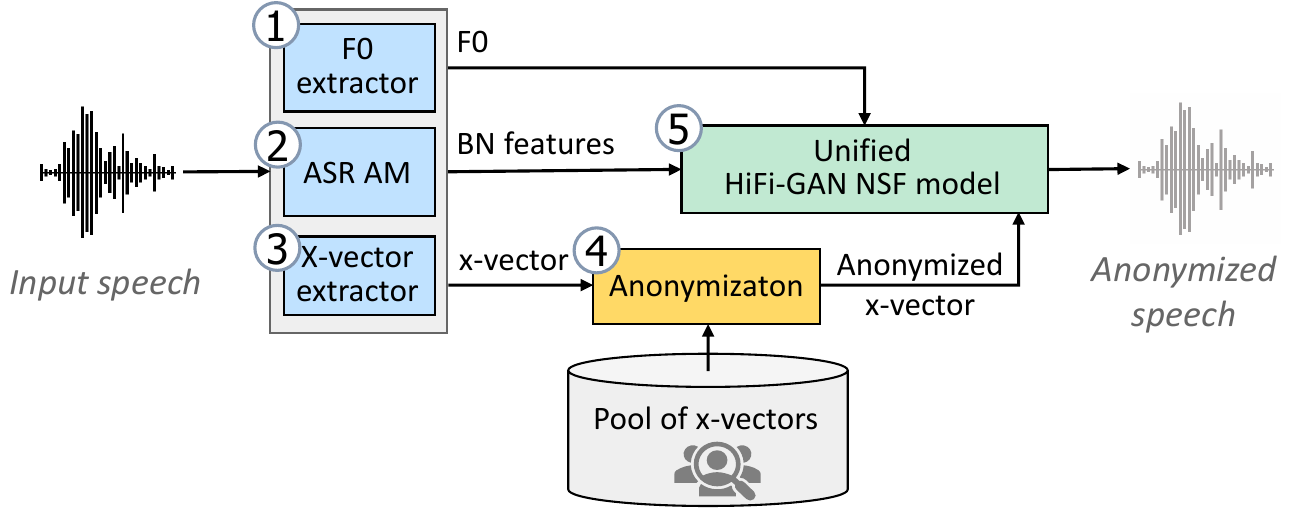}
\caption{Baseline anonymization system \textbf{B1}. 
}
\label{fig:baseline1}
\end{figure}

\textbf{B1} is based on the voice anonymization method proposed in \cite{fang2019speaker} and shown in Figure~\ref{fig:baseline1}.
Anonymization is performed in three steps:
\begin{itemize}
\item \textbf{Step 1 -- Feature extraction:} extraction of the speaker x-vector \cite{snyder2018x}, the fundamental frequency (F0) and 
bottleneck (BN) features
from the original audio waveform.
\item \textbf{Step 2 -- X-vector anonymization:} generation of an anonymized (pseudo-speaker) x-vector using an external pool of speakers.

\item \textbf{Step 3 -- Speech synthesis:} synthesis of an anonymized speech waveform from the
anonymized x-vector and the original BN and F0 features using a neural source-filter (NSF) model.
\end{itemize}

     In order to implement these steps, four different models are required, as shown in Figure~\ref{fig:baseline1}. Details for training these components are presented in Table~\ref{tab:data-baseline-models}.

In \textit{Step 1}, to extract BN features, an ASR acoustic model (AM) is trained (\#1 in Table~\ref{tab:data-baseline-models}). We assume that the BN features represent the linguistic content of the speech signal. The ASR AM has a factorized time delay neural network (TDNN-F)  model architecture \cite{povey2018semi,peddinti2015time} and is trained using the Kaldi toolkit \cite{povey2011kaldi}.
To encode speaker information, an x-vector extractor with a TDNN model topology (\#2 in Table~\ref{tab:data-baseline-models}) is also trained using Kaldi.

In \textit{Step 2}, for a given source speaker, a new anonymized x-vector is computed by averaging a set of candidate x-vectors from the speaker pool. 
Probabilistic linear discriminant analysis (PLDA) is used as a distance measure between these vectors and the x-vector of the source speaker.  
The candidate x-vectors for averaging are chosen in two steps. First, for a given source x-vector, the $N$ farthest x-vector candidates in the speaker pool are selected. Second, a smaller subset of $N^*$ candidates are chosen randomly among those $N$ vectors ($N=200$ and $N^*=100$).
The x-vectors for the speaker pool are extracted from a disjoint dataset (\textit{LibriTTS-train-other-500}).

In \textit{Step 3}, the NSF model used to synthesize the anonymized speech waveform is trained on \textit{LibriTTS-train-clean-100} in the same manner as HiFi-GAN \cite{kong2020hifi} using the HiFi-GAN discriminators. After training, the discriminators can be safely discarded, and only the trained NSF model is used in the anonymization system.

More details about this model can be found in the \href{https://github.com/Voice-Privacy-Challenge/Voice-Privacy-Challenge-2022}{scripts} for VoicePrivacy 2022\footnote{To perform \textit{utterance-level} (in contrast to  \textit{speaker-level}) anonymization of the enrollment and trial data for \textbf{B1}, the corresponding parameters should be setup in \href{https://github.com/Voice-Privacy-Challenge/Voice-Privacy-Challenge-2022/blob/master/baseline/config.sh}{config.sh}: \textcolor{darkspringgreen}{\texttt{anon\_level\_trials=utt}} and \textcolor{darkspringgreen}{\texttt{anon\_level\_enroll=utt}}.}
%\textsuperscript{\ref{fn:scripts}} 
and in \cite{srivastava2020baseline,srivastava2021}.

\begin{table}[h!]
  \caption{Modules and training corpora for the anonymization system \textbf{B1}.  
  The module indexes are the same as in Figure~\ref{fig:baseline1}. Superscript numbers  represent feature dimensions.}\label{tab:data-baseline-models}
 \renewcommand{\tabcolsep}{0.05cm} 
 \renewcommand{\arraystretch}{1.5}
  \centering
%    \resizebox{0.98\textwidth}{!}{
  \begin{tabular}{|c|c|l|l|l|}
\Xhline{0.6pt}
  \textbf{\#} & \textbf{Module} & \textbf{Description} & \makecell[l]{\textbf{Output} \\ \textbf{features}} & \textbf{Data} \\ \hline \hline
  1 & \makecell{F0 \\extractor} &  pYAAPT\protect\footnotemark, uninterpolated  & F0$^{1}$ & - \\ \hline
  2 & \makecell[c]{ASR \\ AM} & \makecell[l]{ TDNN-F\\  Input:  MFCC$^{40}$ + i-vectors$^{100}$ \\ 17 TDNN-F hidden layers \\ Output: 6032 triphone  ids \\ LF-MMI and CE criteria \\ }  & \makecell[l]{ BN$^{256}$ \textrm{ features} \\   extracted from \\ the final hidden \\ layer} &  \makecell[l]{LibriSpeech:\\ train-clean-100 \\ train-other-500} \\ \hline
3 & \makecell{X-vector \\ extractor} & \makecell[l]{TDNN \\ Input: MFCC$^{30}$ \\  7  hidden layers + 1 stats pooling layer \\ Output: 7232 speaker ids \\ CE criterion\\ } & \makecell[l]{ speaker \\ x-vectors$^{512}$} & VoxCeleb-1,2 \\ \hline
  4 & \multicolumn{2}{l|}{  ~~X-vector anonymization module} & \makecell[l]{ 
pseudo-\\speaker \\ x-vectors$^{512}$} &  \makecell[l]{ 
(Pool of \\speakers) \\ LibriTTS:  \\train-other-500} \\ \hline
\makecell[c]{5 \\ 
} & \makecell{NSF \\ model} & \makecell[l]{sinc-hn-NSF in \cite{wang2019neural} + HiFi-GAN discriminators \cite{kong2020hifi} \\ Input: F0$^{1}$+ \textrm{BN}$^{256}$ + \textrm{x-vectors}$^{512}$ \\ Training criterion defined in Hifi-GAN \cite{kong2020hifi} }  & \makecell[l]{ speech waveform} & \makecell[l]{LibriTTS: \\ train-clean-100} \\ \Xhline{0.6pt}
  \end{tabular}%}
\end{table}
\footnotetext{pYAAPT: \url{http://bjbschmitt.github.io/AMFM_decompy/pYAAPT.html}}

\subsection{Anonymization using the McAdams coefficient: B2}

\begin{figure}[htp]
    \centering
    \includegraphics[width=115mm]{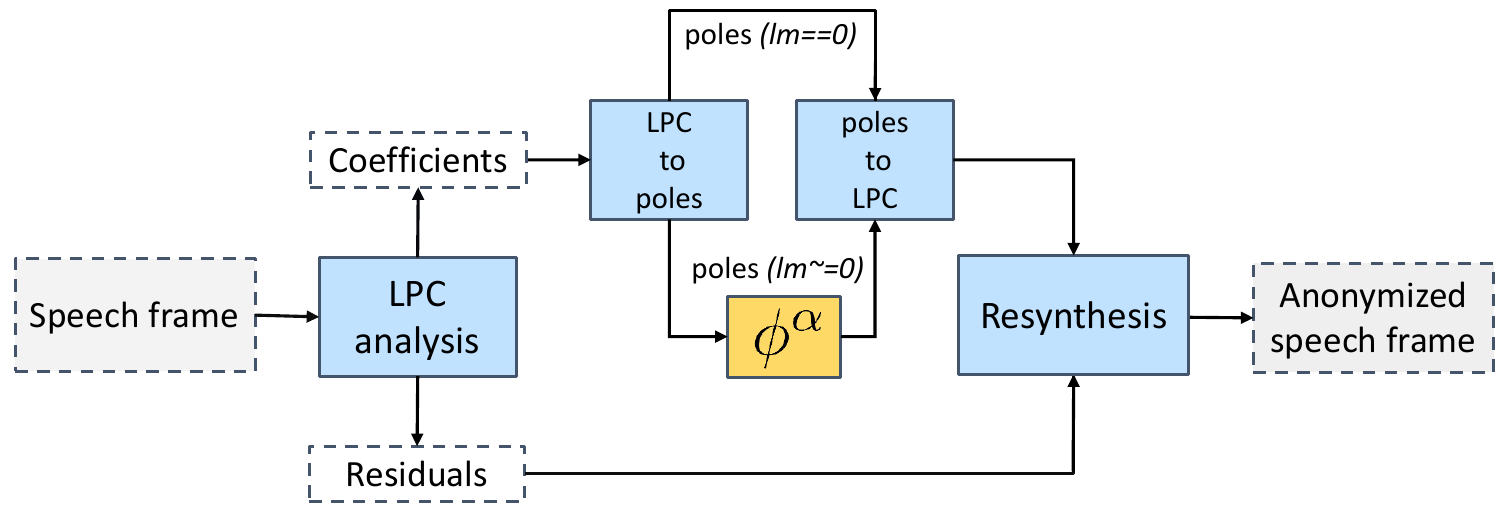}
    \caption{Baseline anonymization system \textbf{B2}. 
    }
    \label{fig:lpc_processing}
\end{figure}

The second baseline anonymization system \textbf{B2} shown in Figure~\ref{fig:lpc_processing} is identical to the \textbf{B2} baseline from the VoicePrivacy 2022 Challenge \cite{tomashenko2022voiceprivacy}.
In contrast to \textbf{B1}, it does not require any training data and is based upon simple signal processing techniques. It is a randomized version of the anonymization method proposed in  \cite{patino2020speaker}, which employs the McAdams coefficient \cite{mcadams1984spectral} to shift the pole positions derived from linear predictive coding (LPC) analysis of speech signals.

\textbf{B2} starts with the application of frame-by-frame LPC source-filter analysis to derive LPC coefficients and residuals. The residuals are set aside for later resynthesis, whereas the LPC coefficients are converted into pole positions in the z-plane by polynomial root-finding. Each pole corresponds to a peak in the spectrum, resembling a formant position.
The McAdams' transformation is applied to the phase of each pole: while real-valued poles are left unmodified, the phase $\phi$ (between 0 and $\pi$~radians) of poles with non-zero imaginary parts is raised to the power of the McAdams' coefficient $\alpha$ so that transformed poles have new, shifted phases of $\phi^\alpha$.
The coefficient $\alpha$ is sampled for each utterance
from a uniform distribution: $\alpha\sim U(\alpha_\text{min},\alpha_\text{max})$, with $\alpha_\text{min}=0.5$ and $\alpha_\text{max}=0.9$. It implies a contraction or expansion of the pole positions around $\phi=1$~radian.
For a sampling rate of 16~kHz, i.e.\ for the data used in this challenge, $\phi=1$~radian corresponds to approximately 2.5~kHz which is the approximate mean formant position \cite{ghorshi2008cross}.
The corresponding complex conjugate poles are similarly shifted in the opposite direction  
and the new set of poles, including original real-valued poles, 
are converted back into LPC coefficients. Finally, the LPC coefficients and the residuals are used to resynthesise a new speech frame in the time domain.

\subsection{Anonymization using  phonetic transcriptions and GAN:
 B3}
\label{sec:baseline_b3}
\begin{figure}[h!]
\centering\includegraphics[width=154mm]{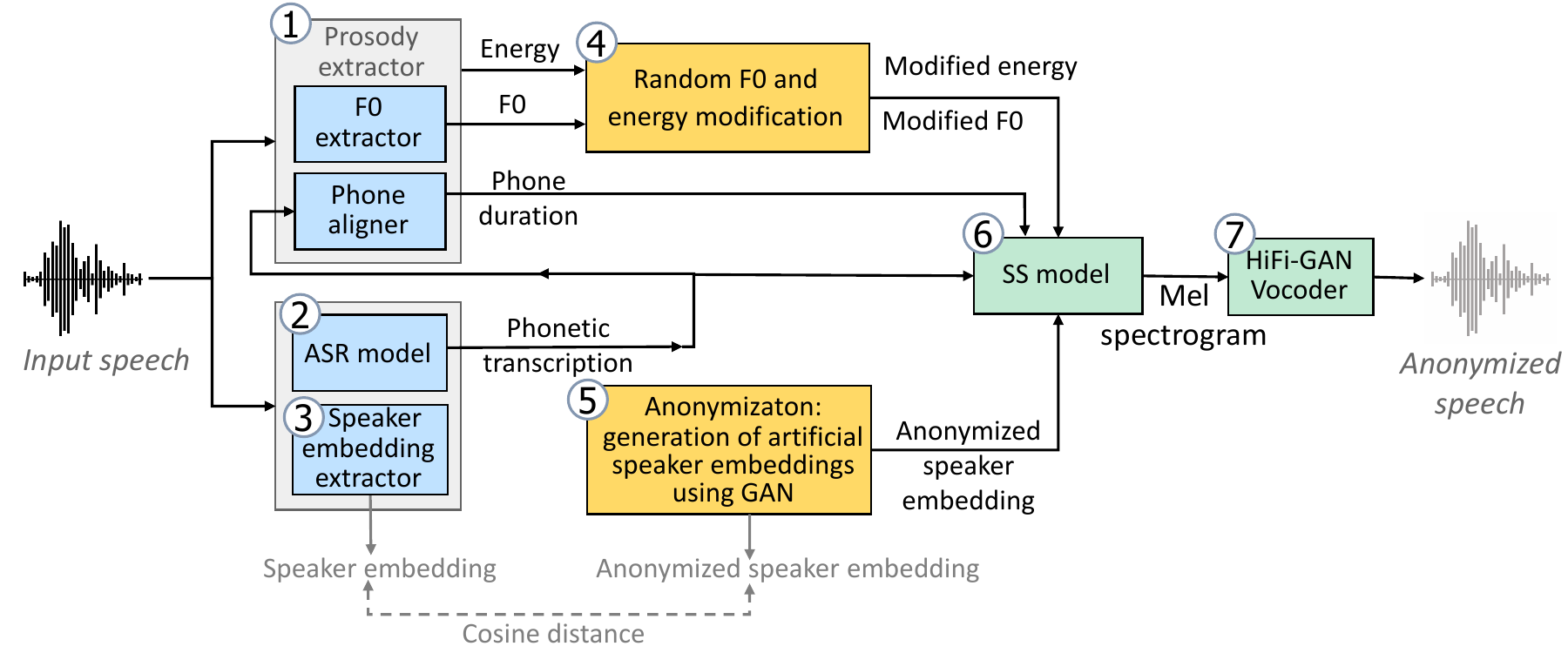}
\caption{Baseline anonymization system \textbf{B3}. 
}
\label{fig:gan_baseline}
\end{figure}

The baseline \textbf{B3}, shown in Figure~\ref{fig:gan_baseline}, is a system based on 
speech synthesis conditioned on keeping linguistic and general prosodic information while replacing the speaker embedding \cite{meyer2023prosody}. 
The core part of the baseline is a generative adversarial network (GAN) that generates artificial pseudo-speaker embeddings \cite{meyer2023anonymizing}. 
Anonymization is performed in 
three steps:
\begin{itemize}
\item \textbf{Step 1 -- Feature extraction:} extraction of the speaker embedding, phonetic transcription, fundamental frequency (F0), energy, and phone duration
from the original audio waveform.
\item \textbf{Step 2 -- Speaker embedding anonymization; pitch and energy modification.} 
\item \textbf{Step 3 -- Speech synthesis:} synthesis of an anonymized speech waveform from the
anonymized speaker embedding, modified F0 and energy features, original phonetic transcripts and original phone durations.
\end{itemize}
Different models are required to implement these steps, as shown in Figure~\ref{fig:gan_baseline}. Details for training these components are presented in Table~\ref{tab:data-baseline-b3}.

In \textit{Step 1}, 
the speaker embedding is extracted using an adapted global style tokens model \cite{wang2018style}.
The phonetic transcription is obtained using an end-to-end ASR model with a hybrid CTC-attention architecture, a Branchformer encoder and a Transformer decoder.

In \textit{Step 2}, the original speaker embedding is replaced by an artificial one generated by a Wasserstein GAN \cite{arjovsky2017wgan}. If the cosine distance between the artificial and the original embeddings exceeds 0.3, they are assumed to be dissimilar enough. Otherwise another artificial embedding is generated until this condition is satisfied. Furthermore, the pitch and energy values of each phone are multiplied by
random values generated uniformly  and independently
between 0.6 and 1.4
to remove individual prosodic patterns while keeping the general prosody of the utterance. The random 
values are
chosen for each phone individually. 

In \textit{Step 3}, the anonymized speaker embedding, modified prosody, and original phonetic transcription are fed into a
speech synthesis system based on \textit{FastSpeech2} \cite{ren2020fastspeech} and HiFi-GAN \cite{kong2020hifi} as implemented in \textit{IMS-Toucan} \cite{lux2021toucan} to synthesize the anonymized speech.

\begin{table}[h!]
  \caption{Modules and training corpora for the anonymization system \textbf{B3}.  
  The module indexes are the same as in Figure~\ref{fig:gan_baseline}. Superscript numbers  represent feature dimensions.}\label{tab:data-baseline-b3}
 \renewcommand{\tabcolsep}{0.05cm} 
 \renewcommand{\arraystretch}{1.5}
  \centering
    \resizebox{\textwidth}{!}{
  \begin{tabular}{|c|c|l|l|l|}
\Xhline{0.6pt}
  \textbf{\#} & \textbf{Module} & \textbf{Description} & \makecell[l]{\textbf{Output} \\ \textbf{features}} & \textbf{Data} \\ \hline \hline
  1 & \makecell{Prosody \\extractor} & \makecell[l]{ Phone aligner: 6-layer CNN + LSTM with CTC loss \\F0 estimation using Praat\\ F0, energy, durations normalized by each vector's mean} & \makecell[l]{ F0$^{1}$, energy$^{1}$ \\ phone durations$^{1}$} & \makecell[l]{LibriTTS: \\ train-clean-100}  \\ \hline
  2 & \makecell[c]{ASR} & \makecell[l]{End-to-end with hybrid CTC-attention \cite{watanabe2017hybrid}\\  Input:  log mel Fbank$^{80}$ \\Encoder: Branchformer \cite{peng2022branchformer}\\Decoder: Transformer \\ Output: phone sequences \\ CTC and attention criteria \\ }  & \makecell[l]{ phonetic \\ transcript \\ with pauses \\ and punctuation} &  \makecell[l]{LibriTTS:\\ train-clean-100 \\ train-other-500} \\ \hline
3 & \makecell{Speaker \\embedding \\ extractor} & \makecell[l]{GST \cite{wang2018style}, trained jointly with SS model \\ Input: mel spectrogram$^{80}$\\  6  hidden layers + 4-head attention \\ Output: GST speaker embedding$^{128}$ } & \makecell[l]{ GST speaker \\ embedding$^{128}$} & \makecell[l]{LibriTTS: \\ train-clean-100} \\ \hline
  4 & \makecell{  ~~Prosody \\ modification \\ module}  & \makecell[l]{Value-wise multiplication of F0 and energy \\with random 
 values  in
  $[0.6,1.4)$
  } 
  & F0$^{1}$, energy$^{1}$ &  - \\ \hline
  5 & \makecell{  ~~Speaker \\ anonymization \\ module}  & \makecell[l]{Wasserstein GAN \\ Input: Random noise$^{16}$ from normal distribution \\ Generator: ResNet with three residual blocks, 150k params \\ Critic: ResNet with three residual blocks, 150k params \\ Output:  \\ MSE and Quadratic Transport Cost \cite{liu2019wasserstein} criteria} & \makecell[l]{ 
pseudo-\\speaker GST \\ embeddings$^{128}$} &  \makecell[l]{ 
LibriTTS:  \\train-clean-100 \\RAVDESS \cite{livingstone2018ryerson} \\ESD \cite{zhou2021seen}} \\ \hline  
\makecell[c]{6 \\ 
} & \makecell{SS \\ model} & \makecell[l]{\textit{IMS Toucan} \cite{lux2021toucan} implementation of \textit{FastSpeech2} \cite{ren2020fastspeech} \\ Input: F0$^{1}$ + energy$^{1}$ + phone duration$^{1}$ \\~~~~~~~~+ phonetic transcript + \textrm{GST embeddings}$^{128}$ \\ Training criterion defined in \textit{FastSpeech2} \cite{ren2020fastspeech} }  & \makecell[l]{mel spectrogram$^{80}$} & \makecell[l]{LibriTTS: \\ train-clean-100} \\ \hline 
\makecell[c]{7 \\ 
} & \makecell{Vocoder} & \makecell[l]{HiFi-GAN vocoder \cite{kong2020hifi} \\ Input: mel spectrogram$^{80}$ \\ Training criterion defined in Hifi-GAN \cite{kong2020hifi} }  & \makecell[l]{ speech waveform} & \makecell[l]{LibriTTS: \\ train-clean-100} \\ \Xhline{0.6pt}
  \end{tabular}}
\end{table}
\footnotetext{pYAAPT: \url{http://bjbschmitt.github.io/AMFM_decompy/pYAAPT.html}}

\subsection{Anonymization using neural audio codec (NAC)  language modeling: B4} %\textcolor{green}{next release v1}}
\begin{figure}[h!]
\centering\includegraphics[height=5cm]{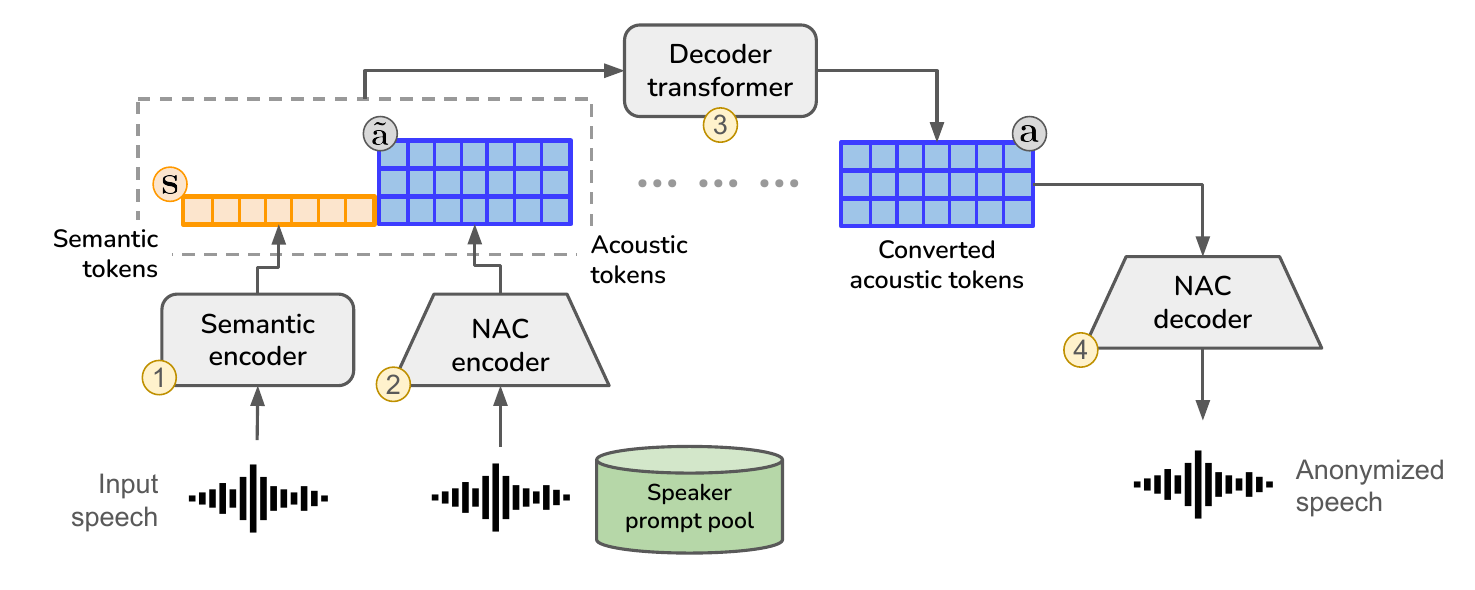}
\caption{Baseline anonymization system \textbf{B4}.}
\label{fig:b4_nac}
\end{figure}

\label{sec:baseline_b4}
System \textbf{B4},  proposed in~\cite{panariello_speaker_2023}, is based upon the technique of neural audio codec (NAC) language modeling~\cite{audiolm, vall-e}. A NAC is an encoder-decoder neural network whereby an audio signal can be encoded into a sequence of discrete \emph{acoustic tokens} $\mathbf{a} \in \{1, \dots, N_Q\}^{Q \times T_A}$ and subsequently decoded to a waveform again;
here, $T_A$ is the number of time frames the input waveform is divided into, and $Q$ is the number of tokens associated to a time frame, each drawn from one of $Q$ different token dictionaries. Tokens are integers ranging from $1$ to $N_Q$.
In the context of \textbf{B4}, acoustic tokens are assumed to capture the characteristics of an individual's speech. Several sets of acoustic tokens are extracted from the speech of a pool of pseudo-speakers, obtaining a \emph{pool of acoustic prompts} $\mathbf{A}$.
Given a speech signal to anonymize, system \textbf{B4} uses a \emph{semantic extractor} to extract from it a sequence of discrete \emph{semantic tokens} $\mathbf{s} \in \{1, \dots, N_S\}^{T_S}$,
where $T_S$ is the number of time frames and $N_S$ is the maximum integer value that a semantic token can take.
The semantic tokens encode the spoken content of the utterance. They are concatenated with a randomly chosen sequence of acoustic tokens $\mathbf{\Tilde{a}} \in \mathbf{A}$ to form a single sequence $(\mathbf{s}, \mathbf{\Tilde{a}})$. A GPT-like, decoder-only Transformer then uses said sequence as a prompt and auto-regressively generates a continuation of acoustic tokens $\mathbf{a}$ that respects both the semantics encoded in $\mathbf{s}$ and the speech style encoded in $\mathbf{\Tilde{a}}$. The decoder module of the NAC is used to convert $\mathbf{a}$ to a waveform that preserves the semantic content of the original input audio, but is associated to a different pseudo-speaker.
An outline of system \textbf{B4} is shown in Figure~\ref{fig:b4_nac}.

The NAC is EnCodec\footnote{\url{https://github.com/facebookresearch/encodec}}\cite{encodec}, which is trained with speech segments from the DNS Challenge~\cite{dubey2022icassp} and \textit{Common Voice}~\cite{ardila2020common}, along with non-speech audio data from \textit{AudioSet}~\cite{audioset}, \textit{FSD50K}~\cite{fsd50k}, and \textit{Jamendo}~\cite{jamendo}.
The semantic extractor is composed of a HuBERT feature extractor~\cite{hubert} trained on \textit{LibriSpeech-train-960}, and a LSTM back-end that predicts a token index from the feature vector at each time frame.
The decoder-only model is a publicly available checkpoint from Bark\footnote{\label{footnote:bark}\url{https://github.com/suno-ai/bark}}, although its authors do not disclose its training data. Further architectural details of the implementation are provided in Table~\ref{tab:data-baseline-b4}.

\begin{table}[h!]
  \caption{Modules and training corpora for anonymization system \textbf{B4}.}\label{tab:data-baseline-b4}
 \renewcommand{\tabcolsep}{0.05cm} 
 \renewcommand{\arraystretch}{1.5}
  \centering
  \begin{tabular}{|c|c|l|l|}
\Xhline{0.6pt}
  \textbf{\#} & \textbf{Module} & \textbf{Description} & \makecell[l]{\textbf{Output} \\ \textbf{features}} 
  \\ \hline \hline
  1 & \makecell{Semantic\\encoder} & \makecell[l]{HuBERT Base
quantized \\ \url{https://dl.fbaipublicfiles.com/hubert/hubert_base_ls960.pt} } & \makecell[l]{1 semantic token\\per time step} 
\\ \hline
  2 & \makecell[c]{NAC encoder} & \makecell[l]{EnCodec 24 KHz~\cite{encodec} encoder \\ \url{https://huggingface.co/facebook/encodec_24khz}}  & \makecell[l]{8 acoustic tokens\\per time step}
  \\ \hline
  3 & \makecell[c]{Decoder\\transformer} & \makecell[l]{Two 12-layer, decoder-only transformers\\operating on different ranges of acoustic tokens\\Taken from Bark\textsuperscript{\ref{footnote:bark}}\cite{panariello_speaker_2023} \\ \url{https://huggingface.co/erogol/bark/tree/main}}  & \makecell[l]{8 acoustic tokens\\per time step} 
  \\ \hline
4 & \makecell{NAC decoder} & \makecell[l]{EnCodec 24 KHz
\cite{encodec} decoder}  & \makecell[l]{ speech waveform} 
\\ \Xhline{0.6pt}
  \end{tabular}%}
\end{table}

\subsection{Anonymization using ASR-BN with vector quantization (VQ): B5 and B6}\label{sec:baseline_b5_b6}
This anonymization pipeline elaborated in \cite{champion2023} shares similarities with \textbf{B1}
and showcases some improvements. In particular, it exclusively relies on PyTorch for execution and has been optimized for fast inference.
Furthermore, it incorporates vector quantization (VQ) to enhance the disentanglement of linguistic and speaker attributes.

\begin{figure}[h!]
\centering\includegraphics[width=120mm]{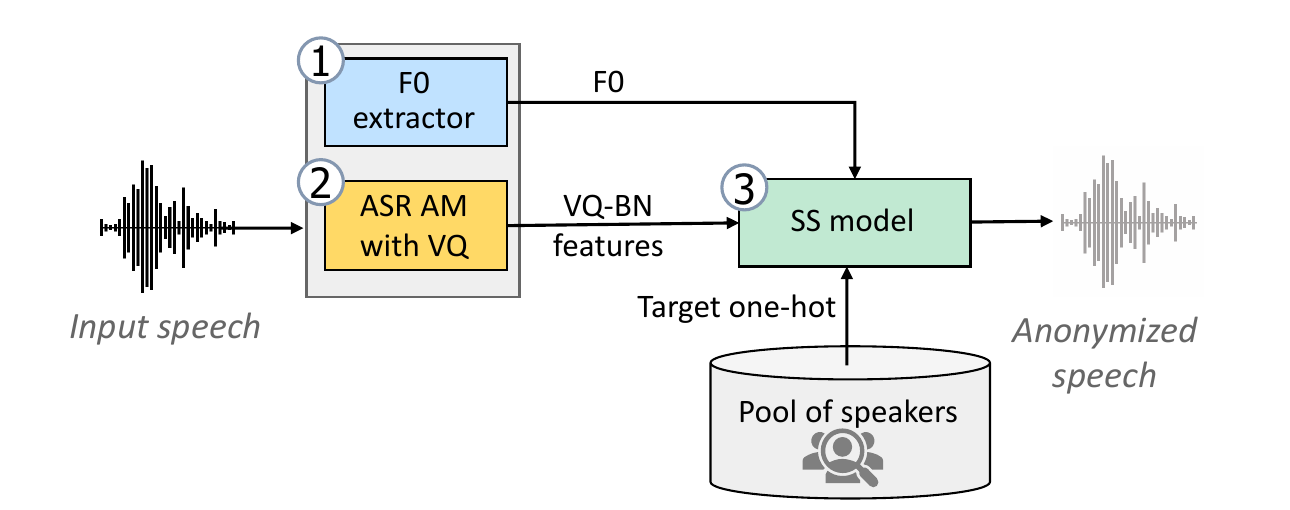}
\caption{Baseline anonymization systems \textbf{B5} and \textbf{B6}.} 
\label{fig:vq_baseline}
\end{figure}

The pipeline leverages feature extractors to capture the fundamental frequency (F0) utilizing a Torch version of YAAPT, and acoustic VQ bottleneck (VQ-BN) features from an ASR AM specifically trained to identify left-biphones.
Subsequently, the VQ-BN features, F0, and a designated speaker (represented as a one-hot vector corresponding to a speaker encountered during training) are directly used to synthesize an anonymized speech waveform via a HiFi-GAN network (see Figure~\ref{fig:vq_baseline}).

We consider two different ASR AMs for feature extraction that correspond to two baselines:
\begin{itemize}[leftmargin=15.5mm]
\item
[\textbf{B5}:] the AM  combines a pretrained wav2vec2 model with three additional TDNN-F layers; 

\item
[\textbf{B6}:] the AM consists solely of 12 TDNN-F layers.
\end{itemize}

On the final TDNN-F layer, following the first activation (inner bottleneck of the TDNN-F with 256 dimensions), vector quantization (VQ) is applied.
This process approximates a continuous vector with another vector of equivalent dimensions, the latter belonging to a finite set of vectors.
The incorporation of VQ into this framework serves the purpose of minimizing the encoding of speaker information within the BN features, thereby enhancing the disentanglement property.

The wav2vec2 model is pre-trained on 24.1k hours of unlabeled multilingual west Germanic speech from \textit{VoxPopuli}\footnote{\url{https://dl.fbaipublicfiles.com/voxpopuli/models/wav2vec2\_large\_west\_germanic\_v2.pt}}, then fine-tuned on \textit{LibriSpeech train-clean-100}.
The TDNN-F-only ASR-BN extractor takes Mel filterbank features as inputs and is trained on \textit{LibriSpeech train-other-500} and \textit{train-clean-100}.
The HiFi-GAN model is trained on \textit{LibriTTS-train-clean-100} for each ASR-BN extractor.
\textit{LibriTTS-train-clean-100} contains 247 speakers, so 247 possible one-hot vectors.

\subsection{Results}\label{subsec:results}

Results for the  baselines are reported in Tables~\ref{tab:asv-results}, \ref{tab:asr-results}, and \ref{tab:ser-results}.
For the two old baselines, we can see that \textbf{B1} achieves a better average EER and WER than \textbf{B2}, while \textbf{B2} achieves a better UAR than \textbf{B1}.
Results for the four new baselines 
\textbf{B3}, \textbf{B4}, \textbf{B5}, and \textbf{B6} show better privacy protection than  the  two old baselines and fall  into two  privacy categories according to the privacy thresholds (Section~\ref{sec:perf_objective}). 
The highest EER is achieved by \textbf{B5}. 
The best result among all the baseline systems
in terms of WER is obtained by \textbf{B1}, and the best result in terms of UAR by \textbf{B2}.

\begin{table*}[b!]
  \caption{EER (\%) achieved on data anonymized by the baselines vs.\ original (Orig.) data.}\label{tab:asv-results}
  \centering
  \begin{tabular}{|l|l||c||r|r|r|r|r|r|}
\Xhline{0.7pt}
 \multirow{2}{*}{	\textbf{Dataset}} &  \multirow{2}{*}{\textbf{Gender}}  & \multicolumn{7}{c|}{\textbf{EER,\%}}\\ \cline{3-9}
&  &{	\textbf{Orig.}} & \multicolumn{1}{|c|}{\textbf{B1}} & \multicolumn{1}{|c|}{\textbf{B2}} &  \multicolumn{1}{|c|}{\textbf{B3}} &  \multicolumn{1}{|c|}{\textbf{B4}} &  \multicolumn{1}{|c|}{\textbf{B5}} &  \multicolumn{1}{|c|}{\textbf{B6}}\\ \hline \hline
\multirow{2}{*}{LibriSpeech-dev}    &   female  &	10.51 & 	10.94 & 	12.91 & 	28.43 & 	34.37 & 	35.82 & 	25.14\\
                                    &	male    &	0.93 & 	    7.45 & 	    2.05 & 	   22.04 & 	   31.06 & 	    32.92 & 	20.96\\ \hline
\multicolumn{2}{|c||}{\cellcolor{gray!7} Average dev} &		\cellcolor{gray!7} 5.72	 & \cellcolor{gray!7}	9.20	& \cellcolor{gray!7} 7.48 	& \cellcolor{gray!7} 25.24 & \cellcolor{gray!7} 32.71 & \cellcolor{gray!7} 34.37 & \cellcolor{gray!7}23.05 \\ \hline\hline
\multirow{2}{*}{LibriSpeech-test}	&	female 		&		8.76  & 	7.47 & 	7.48 & 	27.92 & 	29.37 & 	33.95 & 	21.15\\
	                                &	male 		&		0.42 & 	4.68 & 	1.56 & 	26.72 & 	31.16 & 	34.73 & 	21.14 \\  \hline
 \multicolumn{2}{|c||}{\cellcolor{gray!7} Average test} &		\cellcolor{gray!7}4.59	 & \cellcolor{gray!7}	6.07	 	& \cellcolor{gray!7} 4.52	& \cellcolor{gray!7} 27.32 & \cellcolor{gray!7} 30.26 & \cellcolor{gray!7} 34.34 & \cellcolor{gray!7} 21.14 \\
\Xhline{0.7pt}
  \end{tabular}  
\end{table*}

\begin{table}[h!]
  \caption{WER (\%) achieved on data processed by the baselines vs.\ original (Orig.) data.}\label{tab:asr-results}
  \centering
 \begin{tabular}{|l||r||c|c|c|c|c|c|c|}
\Xhline{0.7pt}
 \multirow{2}{*}{\textbf{Dataset}}    & \multicolumn{7}{c|}{\textbf{WER,\%}}  \\ \cline{2-8}
 &	\textbf{Orig.} & \textbf{B1} & \textbf{B2} & \textbf{B3} &  \textbf{B4} &  \textbf{B5} & \textbf{B6}\\ \hline \hline
LibriSpeech-dev	&	1.80  &  3.07  &  10.44  &  4.29  &  6.15  &  4.73  &  9.69\\ \hline
LibriSpeech-test&	1.85  &  2.91  &  9.95  &  4.35  &  5.90  &  4.37  &  9.09\\ 
\Xhline{0.7pt}
  \end{tabular}  
\end{table}
\normalsize

\begin{table}[h!]
  \caption{UAR (\%) achieved on data processed by the baselines vs.\ original (Orig.) data.}\label{tab:ser-results}
  \centering
 \begin{tabular}{|l||r||c|c|c|c|c|c|c|c|c|}
\Xhline{0.7pt}
 \multirow{2}{*}{\textbf{Dataset}}    & \multicolumn{7}{c|}{\textbf{UAR,\%}}\\ \cline{2-8}
 &	\textbf{Orig.} & \textbf{B1} & \textbf{B2} & \textbf{B3} & \textbf{B4} & \textbf{B5} & \textbf{B6}  \\ \hline \hline
IEMOCAP-dev	&		69.08	&	42.71	&	55.61	&	38.09 & 41.97 & 38.08 & 36.39 \\ \hline
IEMOCAP-test	&	71.06	&	42.78	&	53.49	&	37.57 & 42.78 & 38.17 & 36.13	\\ 
\Xhline{0.7pt}
  \end{tabular}  
\end{table}
\normalsize

\section{Evaluation rules}

\begin{itemize}
    \item Participants are free to develop their own anonymization systems, using components of the baselines or not. These systems must operate on the \textit{utterance level}. 
    \item Participants are strongly encouraged to make multiple submissions corresponding to different privacy-utility tradeoffs (see Section~\ref{sec:asv-eval}).  
     \item The three metrics (EER, WER, UAR) will be used for system ranking on the provided development and evaluation sets. Within each EER interval~-- 
     [10,20), [20,30), [30,40), [40,100)  
     -- systems  will be ranked separately in order of (1) increasing WER and (2) decreasing UAR. 
    \item Participants can use the models and data listed in Table~\ref{tab:data-models-final-list}.
    The use of any other data or models not included in this table is strictly prohibited.
    \item Participants must anonymize the development and evaluation sets and the \textit{LibriSpeech-train-clean-360} dataset used to train the ASV evaluation model using the same anonymization system. They must then train the ASV evaluation model on the anonymized training data and compute the evaluation metrics (EER, WER, UAR) on the development and evaluation sets using the provided scripts. Modifications to the training or evaluation recipes (e.g., changing the ASV model architecture or hyperparameters, retraining the ASR and SER models, etc.) are prohibited.
\end{itemize}

\section{Post-evaluation analysis}\label{sec:posteval}

The organizers will run additional post-evaluation experiments in order to further characterize the performance of submitted systems. To do so, we ask all participants to share with us the anonymized speech data obtained when running their anonymization system on the training, development and evaluation sets. Further details about these experiments will follow in due course.

\section{Registration and submission of results}

\subsection{Registration}

Participants/teams are requested to register for the evaluation.  Registration should be performed \textbf{once only} for each participating entity using the  
\href{https://forms.office.com/r/T2ZHD1p3UD}{registration form}.
Participants will receive a confirmation email within $\sim$24 hours after successful registration, otherwise or in case of any questions they should contact the organizers:
 \begin{center}
 \href{mailto:organisers@lists.voiceprivacychallenge.org?subject=VoicePrivacy 2024 registration}{organisers@lists.voiceprivacychallenge.org}.
\end{center}

Also, for the updates, all participants and everyone  interested the VoicePrivacy Challenge  are encouraged to subscribe to the group:
\begin{center}
\url{https://groups.google.com/g/voiceprivacy}.
\end{center}

\subsection{Submission of results}\label{subsec:submission}

Each participant may submit as many systems as they wish for each minimum target EER provided in Section~\ref{sec:asv-eval}.  
In the case of three or more submissions per condition, the organisers will only include the system with the lowest WER and the system with the highest UAR in the official ranking. These two systems (or this system in case it's the same one) will be ranked in terms of both WER and UAR.

Each single submission should include a compressed archive containing:

\begin{enumerate}
    \item
    Directories with the result files, the corresponding cosine similarity 
 scores (saved in \textcolor{darkspringgreen}{\texttt{exp\slash asv\_orig\slash cosine\_out}} and
\textcolor{darkspringgreen}{\texttt{exp\slash asv\_anon<anon\_data\_suffix>\slash cosine\_out}}), and additional information  generated by the evaluation scripts:

    \begin{itemize}
    \item
        \textcolor{darkspringgreen}{\texttt{exp\slash results\_summary}} 
     \item
        \textcolor{darkspringgreen}{\texttt{exp\slash asv\_orig}}
    \item
        \textcolor{darkspringgreen}{\texttt{exp\slash asv\_anon<anon\_data\_suffix>}}
    \item
        \textcolor{darkspringgreen}{\texttt{exp\slash asr}}
    \item 
        \textcolor{darkspringgreen}{\texttt{exp\slash ser\slash *csv}}.
    
     \end{itemize}

    \item The corresponding anonymized speech data (wav files, 16~kHz, with the same names as in the original corpus) generated from the development and evaluation sets and from the \textit{LibriSpeech-train-clean-360} dataset used to train the ASV evaluation model. 
    For evaluation, the wav files will be converted to 16-bit signed integer PCM format, and this format is recommended for submission.
    These data will be used by the challenge organizers to verify the submitted scores, perform post-evaluation analysis with other metrics and subjective listening tests.
    All anonymized speech data should be submitted in the form of a single compressed archive.

\end{enumerate}

A summary of the WER and UAR results on the development and evaluation sets is saved in a single file \textcolor{darkspringgreen}{\texttt{exp\slash results\_summary}}) \footnote{Example \textit{results} files for the baseline systems:
\begin{itemize}
    \item 
\textbf{B1}:
        \url{https://github.com/Voice-Privacy-Challenge/Voice-Privacy-Challenge-2024/blob/main/results/result_for_rank_b1b}
    \item
\textbf{B2}: \url{https://github.com/Voice-Privacy-Challenge/Voice-Privacy-Challenge-2024/blob/main/results/result_for_rank_mcadams}

    \item 
\textbf{B3}: \url{https://github.com/Voice-Privacy-Challenge/Voice-Privacy-Challenge-2024/blob/main/results/result_for_rank_sttts}

    \item
\textbf{B4}: \url{https://github.com/Voice-Privacy-Challenge/Voice-Privacy-Challenge-2024/blob/main/results/result_for_rank_nac}

    \item
\textbf{B5}: \url{https://github.com/Voice-Privacy-Challenge/Voice-Privacy-Challenge-2024/blob/main/results/result_for_rank_asrbn_hifigan_bn_tdnnf_wav2vec2_vq_48_v1}

    \item
\textbf{B6}: \url{https://github.com/Voice-Privacy-Challenge/Voice-Privacy-Challenge-2024/blob/main/results/result_for_rank_asrbn_hifigan_bn_tdnnf_600h_vq_48_v1}

\end{itemize}
}.

Each participant should also submit a single, detailed system description. 
All submissions should be made according to the schedule below. Submissions received after the deadline will be marked as `late' submissions, without exception.
System descriptions will be made publicly available on the Challenge website.
Further details concerning the submission procedure 
will be published
via \url{https://groups.google.com/g/voiceprivacy}, by email, or 
via the \href{https://www.voiceprivacychallenge.org/}{VoicePrivacy Challenge website}.

\section{VoicePrivacy Challenge workshop at Interspeech 2024}

The VoicePrivacy 2024 Challenge will culminate in a joint workshop held in Kos Island, Greece in conjunction with \href{http://www.interspeech2024.org/}{\textbf{Interspeech 2024}} and in cooperation with the ISCA SPSC Symposium.\textsuperscript{\ref{fn:spsc}}
VoicePrivacy 2024 Challenge participants are encouraged to submit papers describing their challenge entry according to the paper submission schedule
(see Section~\ref{sec:schedule}).
Paper submissions must conform to the format of the ISCA SPSC Symposium proceedings, detailed in the author’s kit\footnote{https://interspeech2024.org/author-resources/}, and be 4 to 6 pages long excluding references. Papers must be submitted via the online paper
submission system. 
Submitted papers will undergo peer review via the regular
ISCA SPSC Symposium review process, though the review criteria applied to regular papers will be adapted for VoicePrivacy Challenge papers to be more in keeping with systems descriptions and results.
Nonetheless, 
the submission of regular scientific papers related to voice privacy and anonymization are also invited and will be subject to the usual review criteria.
%Since subjective evaluation results will be released only after the submission deadline, challenge papers should report only objective evaluation results.
The same paper template should be used for system descriptions but may be 2 to 6 pages in length.

Accepted papers will be
presented at the joint ISCA SPSC Symposium and VoicePrivacy Challenge Workshop and will be published as other symposium proceedings 
in the ISCA Archive. Challenge participants without accepted papers are also invited to participate in the workshop and present their challenge contributions
 reported in system descriptions. 
 
More details will be announced in due course.

\section{Schedule}\label{sec:schedule}

The result submission deadline is \colorbox{cambridgebluemy2}{\textbf{15th June 2024}}.
All participants are invited to present their work at the joint SPSC Symposium and VoicePrivacy Challenge workshop that will be organized in conjunction with Interspeech~2024.

\begin{table}[tbh]
  \caption{Important dates}\label{tab:dates}
  \centering
   \resizebox{\textwidth}{!}{
  \renewcommand{\tabcolsep}{-0.12cm} 
  \begin{tabular}{l r }
    \toprule
  Release of evaluation data, software and baselines & \textcolor{blue}{8th March 2024} \\ \midrule
    Deadline for participants to submit a list for training data and models  & \textcolor{blue}{20th March 2024} \\ \midrule
    Publication of the full final list of training data and models  & \textcolor{blue}{21st March 2024} \\ \midrule
Submission of challenge papers to the joint SPSC Symposium and VoicePrivacy Challenge workshop & \textcolor{blue}{15th June 2024} \\  \midrule
 Deadline for participants to submit objective evaluation results, anonymized data, and system descriptions & \textcolor{blue}{15th June 2024} \\  \midrule 
 Author notification for challenge papers & \textcolor{blue}{5th July 2024} \\  \midrule 
Final  paper upload   & \textcolor{blue}{25th July 2024} \\  \midrule
Joint SPSC Symposium and VoicePrivacy Challenge workshop   & \textcolor{blue}{6th September 2024} 
 \\ \bottomrule
   \end{tabular}
   }
\end{table}

\section{Acknowledgement}
This work was supported by the French National Research Agency under project Speech Privacy and project IPoP of the Cybersecurity PEPR and jointly by 
the French National Research Agency and the Japan Science and Technology Agency under project VoicePersonae. The challenge organizers thank Ünal Ege Gaznepoğlu for his help with the code base.

\bibliographystyle{IEEEtran}
\bibliography{main}

\end{document}